\documentclass[12pt]{article}
\usepackage{amsmath,amssymb}
\usepackage[a4paper,hmarginratio=1:1,vmarginratio=2:3,totalwidth=15.2cm,totalheight=22.6cm]{geometry}
\usepackage{xcolor} 
\usepackage[normalem]{ulem}
\usepackage{cite}
\usepackage{hyperref}

\def\e{\mathcal{E}}

\begin{document}

\thispagestyle{empty}
\vspace{3.cm}
\begin{center}

  {\Large \bf  World-Line Actions in Weyl Geometry}

 \vspace{1.5cm}

{\bf  Cezar Condeescu},\,\,  and  {\bf  Andrei Micu}\,\,
 \footnote[1]{E-mail: ccezar@theory.nipne.ro, amicu@theory.nipne.ro}
 
\bigskip\bigskip

{\small \textit{Department of Theoretical Physics,}
  \smallskip
  
 National Institute of Physics and Nuclear Engineering (IFIN--HH),
 \smallskip
 
 Str. Reactorului 30, M\u{a}gurele, Ilfov, 077125 Romania}
\end{center}

\medskip

\begin{abstract}
    \noindent
    \today\newline
In this paper we construct, from a gauge theory perspective, the world-line action for a particle moving on a time-like curve in Weyl geometry. The action we find
is dimensionless, Weyl invariant, additive and, in general, non-local due to an open Wilson line which we have to add in order to account for a general Weyl field. In special cases, this Wilson line can be local, but the geometry becomes integrable. The action can not be used to measure the proper time as it is dimensionless and no mass parameter is allowed in the symmetric phase of the theory.
We show that the usual conditions for defining proper time: affine parametrization, dimension of time and additivity supplemented by the requirement of Weyl invariance can not be fulfilled simultaneously and therefore no satisfactory notion of proper time exists in the symmetric phase.
Under spontaneous symmetry breaking the particle acquires a mass, the action becomes Riemannian and the proper time can be again defined. We also construct a classically equivalent quadratic action by using an \textit{einbein} on the world-line and show that the non-locality can be seen to arise from integrating out a constrained field.

\end{abstract}

\newpage

\section{Introduction}

Weyl geometry \cite{HW} is an interesting, viable candidate for a quantum theory
of gravity which is both renormalizable\footnote{The price of renormalizability is the potential presence of ghosts. Some solutions have been proposed for conformal gravity \cite{Bender,Maldacena, Mannheim} and quadratic gravity \cite{Kumar} which may be applied to Weyl geometry as well.} \cite{Stelle} and anomaly free \cite{DG1}. It was introduced over a century ago by Weyl as a
unified theory of gravity and electromagnetism, but from the
beginning it was alleged to be inconsistent with the observed world
due to the fact that length of vectors is path dependent. Even after it
was realized that the theory can not describe electromagnetism but rather a scale symmetry, one
debate over the physical viability of the theory still remained, namely whether the theory exhibits a second clock
effect. Only recently it was realized that most arguments against
this theory could be addressed by treating the local symmetry of scale
invariance as a true gauge symmetry. We know now that in the
symmetric phase, parallel transport should be done with
the Weyl covariant derivative \cite{Dirac,DG1,HL1,CGM} which is metric compatible, a derivative which in general can not
be defined by an affine connection. Moreover, physical observables can
only be associated with gauge-invariant quantities and for such
quantities, parallel transport does not give path dependent results. 
Nevertheless, the second clock effect in Weyl geometry has been a subject of long debates
\cite{Perlick, Romero2, Romero3, HL1, Quiros}.
Based on the axiomatic approach of \cite{EPS}, the quantity proposed to be the proper time in Weyl geometry is non-local, even in infinitesimal form, and depends on the history of the path followed by the particle. Then, this non-locality is at the origin of the presumed second clock effect.

We will give a different perspective on the problem and actually argue that no consistent proper time can be defined in Weyl geometry. We have some physical intuition for this: in a scale invariant theory lengths are not observables (only angles) and it is hard to imagine how to measure a dimensionful quantity like the time. Moreover, being an
observable, any definition of the proper time must necessarily be Weyl invariant,
something which was realized only recently \cite{HL1}. Based on \cite{EPS,Perlick,Romero2}, we ask the proper time to be additive, to give an affine parametrization of the curve and to be Weyl invariant. Then, we show that such a quantity with the correct dimension (inverse mass) can not be defined. Specifically, imposing affine parametrization introduces a tension between the dimension of the proper time and it being Weyl invariant. Moreover this tension can not be resolved by adding a compensator field as this spoils the affine parametrization.

Another purpose of our paper is the construction of
a world-line action for a particle moving on a time-like curve in Weyl
geometry.
In a first instance, it was not known whether such an action, whose Euler--Lagrange equations yield the autoparallel equation of Weyl geometry, exists beyond the Weyl-integrable case.\footnote{Conditions for the existence of such actions in theories of gravity with non-metricity were given recently in \cite{LH,Csillag}.} Also the existence of such an action opens up the possibility of studying Weyl geometry at the quantum level using a first-quantized world-line path integral approach \cite{Schubert}.
We are indeed able to construct a Weyl-invariant, dimensionless,
albeit non-local action which correctly reproduces the geodesic
equation in Weyl geometry. The action is also additive, which is
essential for using it in path integral quantization. 
Even if in Riemannian geometry, the world-line action and
the proper time are tightly related, in Weyl geometry we can not use
this dimensionless action to measure the time because the symmetry
does not allow any dimensionful parameter. Breaking the  Weyl symmetry
generates a scale in the form of the vacuum expectation value
(\textit{vev}) of the dilaton field, $\langle \phi \rangle$. In the
vacuum, the world-line action then becomes purely Riemannian, with the
mass parameter generated also by $\langle \phi \rangle$. Therefore,
up to this mass parameter, the world-line action gives us a measure of
the proper time in the broken phase of the theory.

An action for a particle in Weyl geometry which has all the properties listed above was not written before and this constitutes the main novelty of our paper. Parts of it were derived before with the main purpose of finding a definition for the proper time, but it was never emphasized that an action should be dimensionless (in Planck units), from where it becomes clear that such a quantity can not be used to measure the (proper) time. Definitions of an action using a compensator field were also proposed, but, as we show, this leads to the marginal class of Weyl integrable geometries where the Weyl field is not dynamical and the power of having a true gauge theory is lost.
The action we construct is similar to what was used in the literature
as the proper time and which was believed to exhibit a second clock
effect \cite{Romero2}. Therefore, it is not inconceivable that using this action to
measure the proper time in the broken phase of the theory, a second
clock effect appears due to the corrections which Weyl geometry
induces on the Riemannian action. On general grounds, we expect
however that such effects are suppressed by the scale of symmetry
breaking\footnote{Note also that the Weyl gauge field becomes massive and therefore interactions are short ranged.}. Moreover, only particles which couple directly to the
dilaton are prone to exhibit this effect. In the minimal
scenario of coupling Weyl geometry to the standard model this happens for the Higgs field \cite{DG}, and so, even if present, it seems highly non-trivial to test experimentally. Nevertheless, this seems an interesting avenue to
explore, especially in the context of other recent efforts to test the
cosmic imprint of Weyl geometry  \cite{FHNR, DG-inflation,CH,GM-Ligo}.

Finally, by introducing an einbein on the world-line we construct a second version of the action (classically equivalent to the first one) which removes square roots and may be more amenable for path integral quantization. We also argue that the non-locality can be reinterpreted as arising from integrating out a constrained field on the world-line, and therefore a local version of the action is also presented.

The plan of the paper is the following.
We start in section 2 with a presentation of the world-line formalism
in Riemannian geometry. 
In section 3.1 we shortly introduce the essential notions about Weyl
geometry and in section 3.2 we discuss the geodesic equation. In
section 3.3 we derive the world-line action which leads to this
geodesic equation and in section 3.4 we explain the mechanism of
symmetry breaking. Finally, in section 3.5 we discuss the affine
parametrization, which in Riemannian geometry gives the condition for
finding the proper time, and explain why in Weyl geometry this does
not yield a satisfactory definition of proper time. We also discuss here the particle action with world-line einbein.  In section 4 we present our conclusions. 

\section{Preliminaries: general relativity (GR)}

We review some aspects of the propagation of massive particles in GR presented in a form suited for comparisons and generalizations to Weyl gravity. This also fixes some of our notations and conventions. The world-line theory of Einstein's gravity is of course well-known, and for more details we refer the reader to classical textbooks e.g. \cite{Wald,Carroll}.

In GR, the action describing the motion of a massive particle in free fall in a gravitational field described by the metric $g_{\mu \nu}$, is given by\footnote{We consider the signature of the metric to be $(-, +,
  +, +)$.} 
\begin{equation}
  \label{GRaction}
  S = - m \int_\gamma \sqrt {- g_{\mu \nu} \dot x^\mu \dot x^\nu} d \lambda\; ,
\end{equation}
where $m$ is the rest mass of the particle and $\gamma$ is a time-like
curve describing the particle trajectory with tangent vector $\dot
x^\mu = \frac{d}{d\lambda} x^\mu(\lambda)$. We work in natural
  units i.e.~$\hbar=c=1$, where $x^\mu$ has mass dimension $-1$ and
  so, the parameter $m$ also has the role of making the action dimensionless. We are using for the time being an arbitrary parametrization of the
  world-line, $\lambda$, which is considered dimensionless.\footnote{Strictly speaking, in Riemannian geometry, dimensions of arbitrary parametrizations do not play a special role and can be changed at will using the (constant) parameter $m$. In Weyl geometry this is no longer possible and more care is needed.}
The
Euler--Lagrange equations of the action above take the form
\begin{equation}
  \label{GRgeo}
 m g_{ \sigma \rho}\left[ \frac{d}{d \lambda} \left ( \frac {\dot x^\rho}{\sqrt{-g_{\mu \nu}
        \dot x^\mu \dot x^\nu}} \right ) + \mathring \Gamma_{\alpha \beta}^\rho
      \frac{\dot x^\alpha \dot x^\beta}{\sqrt{- g_{\mu \nu} \dot x^\mu
          \dot x^\nu}} \right] = 0\; ,
\end{equation}
where  $\mathring\Gamma^\rho_{\mu \nu}$ are the Christoffel symbols
completely determined by the metric, $\mathring \Gamma_{\mu \nu}^\rho = \tfrac12
g^{\rho \sigma}(\partial_\mu g_{\nu\sigma} + \partial_\nu g_{\mu
  \sigma} - \partial_\sigma g_{\mu \nu})$.
The known form of the geodesic equation is obtained for the normed
vector $u^\rho$
\begin{equation}
  \label{GRgeo2}
  u^\mu \partial_\mu u^\rho + \mathring \Gamma_{\mu \nu}^\rho u^\mu u^\nu
  = 0 \; , ~~\mathrm{where} ~~   u^\rho \equiv \frac{\dot x^\rho}{||\dot
    x||} = \frac{\dot x^\rho}{\sqrt{-g_{\mu \nu} \dot x^\mu \dot x^\nu}}
  \; .
\end{equation}
After expanding equation \eqref{GRgeo} we find the following equivalent form of the equation of motion of a massive particle
\begin{equation}
  \label{GRgeo3}
\frac{m g_{\sigma \rho}}{\sqrt{-g_{\mu \nu} \dot x^\mu \dot x^\nu}} \left[ \ddot x^\rho + \mathring \Gamma_{\alpha \beta}^\rho \dot x^\alpha \dot
  x^\beta -  \frac{d}{d \lambda} \left( \ln{\sqrt{- g_{\alpha \beta} \dot x^\alpha
          \dot x^\beta}} \right) \dot x^\rho\right] =0 \; .
\end{equation}
After dropping the prefactor above (the metric assumed non-degenerate) one obtains the geodesic equation in an arbitrary parametrization (sometimes also called pregeodesic)
\begin{equation}
\dot x^\mu \mathring \nabla_\mu \dot x^\rho = f(\lambda) \dot x^\rho\; ,
\label{GRpregeo}
\end{equation}
where  $f= \frac{d}{d\lambda} \ln ||\dot x||$ and we used
  $\tfrac{d}{d \lambda} = \dot x^\nu \partial_\nu$. In general, for an
  arbitrary function $f(\lambda)$, 
equation \eqref{GRpregeo} is said to define a projective
structure \cite{EPS,SMZ}.\footnote{Intuitively a geodesic is a generalization of a
straight line in curved geometry. Infinitesimally around a given
point, the solutions of equation \eqref{GRpregeo} are straight lines passing
through that point similar to the definition of a real projective
space.} Two connections belong to the same projective
structure if they have the same (pre)geodesics. In this paper we
  shall consider only functions of the form $f(\lambda) = \dot x^\mu
v_\mu(x(\lambda))$ with $v_\mu$ a spacetime vector. In
  particular, for equation \eqref{GRpregeo} we have $v_\mu =\partial_\mu
\ln ||\dot x||$.
Then a projectively equivalent connection of $\mathring \Gamma$ is 
\begin{equation}
\mathring \Gamma^\rho_{\mu \nu} \rightarrow \check \Gamma_{\mu \nu}^\rho = \mathring   \Gamma^\rho_{\mu \nu} - \delta^\rho_\nu v_\mu\; .
\end{equation}
In terms of the new connection $\check \Gamma$ (and its associated covariant derivative $\check \nabla$) the (pre)geodesic equation becomes
\begin{equation}
\dot x^\mu \check \nabla_\mu \dot x^\nu = 0 \; .
\end{equation}
The form above is called an affine parametrization (w.r.t. $\check
\Gamma$) of the geodesic equation. One can also find an affine
parametrization with respect to the Levi-Civita connection $\mathring
\Gamma$. This is defined as a reparametrization of the world-line
$\lambda \rightarrow \tilde \tau(\lambda)$ such that one has
\footnote{Explicitly, under such a reparametrization $f(\lambda)
  \rightarrow f(\lambda) \dot{\tilde \tau} - \ddot{\tilde \tau} $.}
\begin{equation}
  f(\tilde\tau) = 0\; .
\end{equation}
As can be seen from equation \eqref{GRgeo3}, this can be achieved by imposing
\begin{equation}
  \label{prtcond}
  g_{\mu \nu} \frac{ dx^\mu}{d \tilde\tau} \frac{d x^\nu}{d \tilde\tau} = - k^2 \; ,
\end{equation}
for some constant $k$ with dimension of inverse mass, which also has the role of making this relation dimensionally consistent ($\tilde \tau$ is still assumed dimensionless). In this parametrization, the geodesic equation reads 
\begin{equation}
 \dot x^\mu \mathring \nabla_\mu \dot x^\rho = 0 \; ,
\label{GRgeo4}
\end{equation}
where now the dot denotes the derivatives with respect to
$\tilde \tau$. This form of the
geodesic equation is actually preserved by affine reparametrizations 
\begin{equation}
  \tilde \tau \rightarrow a \tilde \tau + b \; ,
  \label{freedom}
\end{equation}
with $a,b$, real constants. We can thus define $\tau = k
  \tilde\tau$ such that equation \eqref{prtcond} becomes the usual definition
  of the proper time
\begin{equation}
  \label{GR-pt}
  d\tau^2 = - ds^2 = - g_{\mu \nu} dx^\mu dx^\nu \; ,
\end{equation}
with the particle action written as
\begin{equation}
S = - m \int d\tau \; .
\label{pt-action}
\end{equation}

In many situations, and especially in path integral quantization, an
action given in terms of a square root is not easily manageable. Furthermore, equation \eqref{GRaction} or equation \eqref{pt-action} can not be used for massless particles whose motion is described by light-like curves. We can write a classically equivalent action, quadratic in the $\dot x^\mu$, by introducing an einbein $\e$ on the world-line\footnote{One can interpret the action as a one-dimensional sigma model with kinetic term and cosmological constant term containing the mass.}
\begin{equation}
  \label{Se}
  S = \frac12 \int_\gamma \left ( \e^{-1} g_{\mu \nu} \dot x^\mu \dot
    x^\nu -m^2 \e\right ) d \lambda\; .
\end{equation}
Indeed, after eliminating $\e$, assumed positive as a volume form, from its equation of motion
\begin{align}
  \frac{-g_{\mu \nu} \dot x^\mu \dot x^\nu}{\e^2} \equiv \frac{||\dot x ||^2}{\e^2} =  m^2 \; , && \text{therefore}&& \e = \frac{||\dot x||}{m} \; ,
\end{align}
we obtain again equation \eqref{GRaction}. 
 The action \eqref{Se} is reparametrization invariant
$\lambda \to \lambda'(\lambda)$, 
provided $\e$ transforms like a world-line einbein , that is $\e \to \e' = \e \tfrac{d \lambda}{d \lambda'}$. The equation of motion of the particle coordinate $x^\mu(\lambda)$ reads
\begin{equation}
\e^{-1} g_{\sigma \rho} \left[\ddot x^\rho + \mathring \Gamma^\rho_{\alpha \beta} \dot x^\alpha \dot x^\beta - \left(\frac{d}{d\lambda}\log \e\right)\dot x^\rho\right] = 0 \; .
\end{equation}
It can also be written compactly as (after multiplying with the inverse metric)
\begin{equation}
\dot x^\mu \mathring \nabla_\mu \left(\e^{-1} \dot x^\rho\right)=0 \; .
\end{equation}
The forms above suggest that choosing the einbein to be constant, via a reparametrization of the world-line, yields again an affine parametrization. In our conventions (see table \ref{dimensions}), $\e$ has dimensions of inverse mass squared therefore the most natural constant gauge fixing is
\begin{equation}
\e = \frac{1}{m^2} \; .
\label{gauge-e}
\end{equation}
Replacing the above in the equation of motion of the einbein one
obtains the condition \eqref{prtcond} with $k=1/m$.
After rescaling with the mass, $\tau = m^{-1} \tilde \tau$, using the
freedom \eqref{freedom}, this yields the usual GR definition of proper time,  
equation \eqref{GR-pt}. Hence the action in the proper time parametrization becomes
\begin{equation}
  \label{action-R}
  S = \frac{m^2}2 \int \left(g_{\mu \nu} \frac{ dx^\mu}{d\tilde \tau}  \frac{ dx^\nu}{d \tilde \tau} - 1\right) d \tilde \tau \sim \frac{m}2 \int g_{\mu \nu} \frac{d x^\mu}{d\tau} \frac{d x^\nu}{d\tau}  d \tau  \; ,
\end{equation}
where in the gauge fixed form of the action one can drop the constant since it does not contribute to the equations of motion.
The Euler-Lagrange equations yield immediately the geodesic
equation \eqref{GRgeo4} in an affine parametrization (where the derivatives are taken with respect to the proper time $\tau$).
Finally let us notice that the geodesic equation \eqref{GRgeo2}
 can be written in a more concise way (known as autoparallel curves)
\begin{equation}
  \label{Rgeo-cov}
  u^\mu \mathring \nabla_\mu u^\rho = 0 \; ,
\end{equation}
which is useful for generalizing to Weyl geometry. From above one can find again an affine parametrization by simply imposing
\begin{equation}
u^\mu = \dot x^\mu \; ,
\end{equation} 
which is equivalent to equation \eqref{prtcond}.

As a conclusion to this section, we see from equation \eqref{pt-action}
that the action from which the geodesic equation in an affine
parametrization can be derived is the same as the proper time,
up to the mass parameter $m$. 
For the equations of motion, this
mass parameter is completely irrelevant as it is just a multiplicative
constant factor. We formally need it in order to make the action
dimensionless (in units of $\hbar$). 
In the Weyl geometry case which
we treat below, mass parameters, as well as any dimensionful
parameters, are not allowed by the scale symmetry. Therefore, even if
we are able to construct an action from which the geodesic equation
can be derived, this does not automatically give us a notion of proper
time. Moreover, affine parametrizations in Weyl gravity can acquire
the interpretation of proper time only after scale symmetry breaking.

\section{World-line action in Weyl geometry}

\subsection{Weyl geometry}

Weyl geometry (or Weyl quadratic gravity) can be constructed as the gauge theory of scale transformations (also called dilatations). The field
content of the theory comprises the metric
$g_{\mu \nu}$\footnote{Actually, the proper gauging procedure on the tangent
  space makes use of the vielbein $e_\mu^a$ \cite{CGM,CM}, but for the present purposes there is no advantage in using the vielbein.} and the Weyl gauge field
$b_\mu$. The spacetime Lagrangian of the theory is quadratic in the associated curvatures.  Therefore, the theory propagates a scalar mode $\phi$
coming from the $R^2$ term contained in the action
\cite{DG0}. Specifically it is given by
\begin{equation}
  \label{phidef}
  \phi^2 = -R \; ,
\end{equation}
where $R$ is the scale covariant curvature scalar of Weyl gravity \cite{CGM}.
All these will be considered background fields for the propagation of time-like particles, as it was
the case with the metric in Riemannian geometry and will not be varied
in the world-line action. We nevertheless retain their scaling properties and
consider the metric to have charge $+2$ and the dilaton $\phi$ to have
charge $-1$ under Weyl transformations.\footnote{In the minimal
  coupling scenario, once the charge of the metric is fixed, the other
  charges can be unambiguously defined.} Explicitly their
transformations read
\begin{equation}
  \label{gphitr}
  g_{\mu \nu} \to e^{2\Sigma} g_{\mu \nu} \; , ~~~~~~ \phi \to
  e^{-\Sigma} \phi \; ,
\end{equation}
while for the gauge field we have
\begin{equation}
  \label{btr}
  b_\mu \to b_\mu - \partial_\mu \Sigma \; .
\end{equation}

Historically, Weyl geometry was described in terms of a symmetric (torsionless)
connection which is non-metric 
\begin{equation}
  \label{tgamma}
  \tilde \Gamma^\rho_{\mu \nu} = \mathring \Gamma^\rho_{\mu \nu} +
  b_\mu \delta_\nu^\rho + b_\nu \delta_\mu^\rho - g^{\rho
    \sigma}g_{\mu \nu} b_\sigma \; .
\end{equation}
More recently, constructing the theory on the tangent space, it was
realized that this old approach is completely equivalent with the
description in terms of a metric connection with torsion \cite{CGM,CM}
\begin{equation}
  \label{gamma}
  \Gamma^\rho_{\mu \nu} = \mathring \Gamma^\rho_{\mu \nu} +
  b_\nu \delta_\mu^\rho - g^{\rho
    \sigma}g_{\mu \nu} b_\sigma \; ,
\end{equation}
and the two connections are related by a projective transformation
\begin{equation}
  \label{ptr}
  \tilde \Gamma^\rho_{\mu \nu} = \Gamma^\rho_{\mu \nu} +
  \delta_\nu^\rho b_\mu \; . 
\end{equation}
Thus non-metricity is not a feature of Weyl geometry, but rather a choice of connection which is not necessarily physical. In the same way the torsion connection gives a
different perspective on the theory.
Due to the non-geometric origin of the Weyl charges none of these
connections can give a complete description of the theory. Rather we
need a non-affine gauge covariant derivative, $\hat \nabla$, which properly
takes into account these charges.
Its action on a $(r,p)$-tensor with components $X^{\mu_1 \ldots \mu_r}{}_{\nu_1
  \ldots \nu_p}$ is
\begin{equation}
  \hat \nabla_\mu X = \nabla_\mu X + q_X b_\mu X = \tilde \nabla_\mu X
  + \tilde q_X b_\mu X \; ,
  \label{action-gauge}
\end{equation}
where $\nabla$ and $\tilde \nabla$ are the (affine) covariant
derivatives with the connection $\Gamma$ and $\tilde \Gamma$
respectively. $\tilde q_X$ represents the spacetime Weyl charge 
while $q_X$ is the (intrinsic) tangent space charge and the two are
related by
\begin{equation}
  q_X = \tilde q_X + r- p \; ,
  \label{charges}
\end{equation}
obtained easily by using the vielbein in converting indices from
spacetime to tangent space. We see that only in the particular cases of zero
tangent space charge, $q=0$, or zero spacetime charge, $\tilde q=0$,
the action of $\nabla$ and $\tilde \nabla$ respectively yields again a
scale covariant object.

Finally, in
natural units, the mass
dimension for the dilaton $\phi$ and for the Weyl field $b_\mu$ is
$+1$, while the metric has vanishing mass dimension (see table \ref{dimensions}).

\subsection{Geodesic equation in Weyl geometry}

Now we want to generalize the results in section 2 to Weyl geometry.
In GR we started from an action \eqref{GRaction} giving the geodesic
equation in an arbitrary parametrization, equation \eqref{GRgeo4}, and then found
the proper time among the affine parametrizations. In Weyl geometry,
as we shall see, we can not define a Weyl-invariant world-line parameter
$\tau$, with dimension of time, which provides at the same time an affine
parametrization of the geodesic equation. On the other hand we can
define a Weyl invariant dimensionless action which gives rise to the geodesic 
equation. Strictly speaking we are considering autoparallel curves in
Weyl gravity, the natural generalization of the geodesics in GR, with the defining equation \cite{HL1,GM,GM-Ligo}
\begin{equation}
  \label{Wgeo-cov}
  u^\mu \hat \nabla_\mu u^\rho =0 \; .
\end{equation}
Compared to equation \eqref{Rgeo-cov}, we simply replaced  the affine covariant derivative $\mathring \nabla$ by the fully Weyl gauge covariant
derivative \eqref{action-gauge}.
We shall first work with $u^\mu$ defined in equation \eqref{GRgeo2} and later
we shall explain how to identify it with $\dot x^\mu$. A good reason
to work with $u^\mu$ of equation \eqref{GRgeo2} is that it has Weyl charge
$-1$ irrespective of the charge of $\dot x^\mu$. On the tangent space
$u^\mu$ has vanishing Weyl charge and therefore is a geometric vector
and the equation above expands as
\begin{equation}
  \label{Wgeo-u}
  u^\mu \hat \nabla_\mu u^\rho = u^\mu \nabla_\mu 
  u^\rho = u^\mu \partial_\mu  u^\rho + u^\mu \Gamma_{\mu \nu}^\rho
  u^\nu = 0 \; ,
\end{equation}
with the metric connection with vectorial torsion $\Gamma$ defined in equation \eqref{gamma}. The same equation can be
written in terms of the torsionless non-metric connection $\tilde \Gamma$
\eqref{tgamma}
\begin{equation}
  u^\mu \partial_\mu  u^\rho + u^\mu \tilde \Gamma_{\mu \nu}^\rho
  u^\nu - u^\mu b_\mu u^\rho = 0  \; ,
\end{equation}
or, after defining $f:=u^\mu b_\mu$,
\begin{equation}
u^\mu \tilde \nabla_\mu u^\rho= f u^\rho \; .
\end{equation}
These equations are of the form \eqref{GRpregeo}
which means that the covariant derivatives $\hat \nabla$, $\tilde \nabla$ and $\nabla$ belong to the same projective structure, i.e. we can interpret one as a different (in general non-affine) parametrization of the other two. In this sense the 
propagation of particles in Weyl geometry also exhibits an equivalence
between non-metricity and torsion\footnote{Notice that the affine parametrizations of the covariant derivatives are different.}. At the level of the spacetime quadratic action, the  projective transformation \eqref{ptr} is a symmetry \cite{CGM}. Moreover, it has also been shown that it is built-in Weyl geometry as it corresponds to a redefinition of generators of the gauge algebra of the Weyl group (defined as Poincar\'e extended with scale transformations as a semi-direct product) \cite{CM}.

\subsection{World-line action for a particle in Weyl geometry}

We now want to find an action from which equation \eqref{Wgeo-u} can be
derived. There are a few principles we should follow in finding the
action 
\begin{enumerate}
\item it should be Weyl invariant,
\item it should be dimensionless,
\item in an appropriate limit it should reproduce the Riemannian case.
\end{enumerate}
The action we had in GR is not suitable for Weyl
  geometry. First of all, it contains a mass parameter which now is
  forbidden by scale symmetry. Second of all, the action is also not
  Weyl invariant. To see this, note that the coordinates are not
  changed under Weyl transformations and so, a charge of $\dot x^\mu$
  can appear only through the parametrization $\lambda$. Therefore we
consider
\begin{equation}
  \label{charges}
  \left[\dot x^\mu \right ]_W = -   \left[ d\lambda \right ]_W = p -
  ~\mathrm{arbitrary} \; ,
\end{equation}
and so, the action \eqref{GRaction} has Weyl charge $+1$, given solely
by the metric, independent of $p$.
For convenience, we summarize in table \ref{dimensions} the charges of different
objects we use.
\begin{table}[h]
  \centering
  \begin{tabular}[h]{c|c|c|c|c|c|c}
    & $\dot x^\mu$ & $d \lambda$ & $g_{\mu\nu}$ & $\phi$ & $b_\mu$ &
                                                                $W(\lambda)$
   \\
    \hline \hline
    Weyl charge & $p$ & $-p$ & $+2$ & $-1$ & -- & $-1$\\
    \hline
    mass dimension & $-1$ & $0$ & $0$ & $1$ & $1$ & $1$  
  \end{tabular}
  \caption{Weyl charges and mass dimensions}
  \label{dimensions}
\end{table}

\subsubsection{Integrable Weyl geometry}

Even though the GR action
is not invariant, it does transform covariantly. Therefore, an
immediate guess to solve the problems mentioned above is to use a (scalar) compensator field $\phi$
\begin{equation}
  \label{action-IW1}
  S= -\xi \int_\gamma \phi \sqrt{- g_{\mu \nu} \dot x^\mu \dot x^\nu} d
  \lambda \; .
\end{equation}
In order to make the action Weyl invariant and dimensionless this
field must have mass dimension $+1$ and Weyl charge $-1$. 
As we anticipated in equation \eqref{action-IW1}, the dilaton $\phi$, introduced before in equation \eqref{phidef}, has precisely
these properties as can also be seen from table \ref{dimensions}.
$\xi$ is a dimensionless (constant) coupling which gives the
strength of the interaction of the particle with the background
dilaton field and ultimately, the mass of the particle after symmetry breaking.\footnote{Particles can remain massless after the
  breakdown of the scale symmetry, provided $\xi=0$. In the (minimal)
  coupling of Weyl gravity to the Standard Model, the Higgs field
couples directly to the dilaton and acquires a mass upon Weyl
symmetry breaking. Under certain conditions, this can generate a
tachyonic potential for the Higgs field which further triggers
the electro-weak symmetry breaking which generates masses for the
fermions and for the $W^\pm$ and $Z$ bosons \cite{DG}.}

It is straightforward to derive the Euler-Lagrange equations
corresponding to the action \eqref{action-IW1}
\begin{equation}
  \frac{d}{d \lambda} \left(\frac{\phi\, g_{\mu \nu} \dot x^\nu}{\sqrt
      {-\dot x^2}}\right) + \partial_\mu \phi \sqrt {-\dot x^2} -
  \frac12 \frac{\phi\, \partial_\mu g_{\rho \sigma} \dot x^\rho \dot
    x^\sigma}{\sqrt{-\dot x^2}} =0 \; ,
\end{equation}
where we have dropped the overall constant $\xi$.
The terms without derivatives on $\phi$ group themselves in the same
way as before. Introducing again $u^\mu$ from equation \eqref{GRgeo2} and further dividing by $\phi$ and multiplying by the inverse metric, we find
\begin{equation}
 \frac{d}{d\lambda} u^\rho +
    \dot x^\mu \mathring \Gamma_{\mu \nu}^\rho u^\nu + \dot x^\mu
    \left( \delta_\mu^\rho \frac1\phi \partial_\nu \phi -
      \frac{g_{\mu \nu}g^{\sigma \rho}}\phi \partial_\sigma \phi
    \right) u^\nu =0 \; .
\end{equation}
Using $\tfrac{d}{d\lambda} = \dot x^\mu \partial_\mu $ and
comparing to equation \eqref{gamma} we see that this equation can be written in the
form of equation \eqref{Wgeo-u} 
\begin{equation}
  u^\mu \partial_\mu u^\rho + \Gamma^\rho_{\mu \nu} u^\mu u^\nu = 0 \; ,
\end{equation}
with the identification 
\begin{equation}
b_\mu = \frac1\phi \partial_\mu \phi\; .
\label{pure-gauge}
\end{equation}
Therefore, if $\phi$ is local, the action \eqref{action-IW1}
corresponds to an integrable Weyl geometry\footnote{{\it Apriori} the gauge of $b_\mu $ does not have to be the dilaton $\phi$. However the equation of motion of the gauge field imposes \eqref{pure-gauge} if one sets the field strength to zero \cite{CM}. } which actually gives the known
action of a conformally coupled scalar (see e.g.~\cite{Modesto}). This is not surprising given the equivalence between integrable Weyl geometry and conformal geometry \cite{CM}. 

Finally, an interesting aspect about the action \eqref{action-IW1} is that
it tells us what happens under scale symmetry breaking, when the dilaton $\phi$ acquires a vacuum expectation value: the vev of $\phi$ combines with the coupling $\xi$ to produce the
mass of the particle, $m = \xi \langle \phi \rangle$, and the action becomes that of a massive particle in Riemannian geometry.

\subsubsection{Non-integrable Weyl geometry}

The lesson we learned from the previous subsection is that adding
a nowhere vanishing scalar field $\phi$ to the action
\eqref{action-IW1} yields the geodesic equation of Weyl 
integrable geometry with the Weyl field given by equation \eqref{pure-gauge}. 
Reversing the arguments, given a completely general gauge field
$b_\mu$, we can look for a solution to equation \eqref{pure-gauge},
which inserted in the action will give the geodesic equation in Weyl
geometry for the general Weyl field we started with.
Notice that using the gauge covariant derivative
\eqref{action-gauge} and the Weyl charge of $\phi$, equation \eqref{gphitr},
this equation can be written as 
\begin{equation}
\hat \nabla_\mu \phi = 0 \; .
\label{cov-constant}
\end{equation}
{As explained above, the solution to this equation in the
  non-integrable case has to be non-local and so, we look for a
  functional
\begin{equation}
  W(\lambda) = W(b(\lambda)) \; ,
\end{equation}
which depends explicitly on the Weyl gauge field and which transforms
covariantly as
\begin{equation}
  W(\lambda) \rightarrow e^{-\Sigma} W(\lambda) \; , 
  \label{scale-W}
\end{equation}
satisfying also the differential equation \eqref{pure-gauge} with $\phi$ replaced by $W$ 
\begin{equation}
  \label{eq-W}
 \hat \nabla_\mu W \equiv \partial_\mu W - b_\mu W =0 \; .
\end{equation}
Then, the action would be of the form
\begin{equation}
  \label{action-W}
  S = -\xi \int W(\lambda) \sqrt {-g_{\mu \nu} \dot x^\mu \dot x^\nu} d \lambda  \; ,
\end{equation}
and it would be scale invariant. Such a candidate W is given by an open Wilson line which is a solution to
the following equation
\begin{equation}
  \label{Wilson}
  \hat D_\lambda W(\lambda) = \frac{d}{d\lambda} W(\lambda) -
  b(\lambda) W(\lambda) = 0 \; ,
\end{equation}
depending on the Weyl gauge field induced on the world-line, 
  $b(\lambda):= \dot x^\mu b_\mu$, as $W$ has non-vanishing charge
(here considered $-1$). The gauge covariant derivative above is a pull-back on the world-line of $\hat \nabla$. Explicitly it reads
\begin{equation}
\hat D_\lambda = \dot x^\mu \hat \nabla_\mu \; ,
\end{equation}
and therefore equation \eqref{Wilson} is simply a rewriting of the condition
\eqref{eq-W}. 
The covariant constancy condition \eqref{Wilson} on the world-line is a first order differential equation in one variable which can formally be integrated to 
\begin{equation}
  \ln \frac{W(\lambda)}{W(\lambda_0)} = \int_{\lambda_0}^{\lambda} b(\lambda') d\lambda' \; ,
\end{equation}
and we have the final solution of equation \eqref{Wilson} as
\begin{equation}
  \label{Wsol}
  W(\lambda) = W(\lambda_0) e^{\int_{\lambda_0}^\lambda  b(\lambda') d\lambda'} = 
    W(\lambda_0) e^{\int_{x(\lambda_0)}^{x(\lambda)} b_\mu dx^\mu} \; .
\end{equation}
Here, $\lambda_0$ denotes the starting point on the path followed by
the particle. For the moment we do not specify it, but it can in
principle be $- \infty$ or any point on the path, from where we decide
to follow the particle. $W(\lambda_0)$ is an integration constant which does not
depend on $\lambda$.

The derivation of the Euler--Lagrange equations is very similar to the
pure gauge case since $W$ satisfies equation \eqref{eq-W} (equivalent for this purpose to equation \eqref{pure-gauge} for $\phi$) which immediately leads to equation \eqref{Wgeo-u} for a general
Weyl field $b_\mu$. The fact that now we do not have a Weyl integrable
geometry comes at the cost that the factor $W(\lambda)$ which we added
to the action is non-local, meaning that it depends on the history of the path followed
by the particle. This is in agreement with what we explained before,
that a local factor $W(\lambda) = \phi(\lambda)$ leads to a geodesic
equation with a Weyl field which is pure gauge.

Strictly speaking, in order to obtain covariant equations of motion,
one does not need the action to be invariant (e.g. invariance up to a
constant global factor could be allowed or up to boundary terms like
the Lorentz force in electrodynamics).\footnote{Although not Weyl-invariant, such a Lorentz coupling to non-metricity was presented recently in \cite{DI}.} However, for the purposes of
path integral quantization it is desirable to have an invariant
action. As we already anticipated in table \ref{dimensions},
this requirement, together with the fact that the action has to be
dimensionless in natural units implies that the ``constant" in front of
equation \eqref{Wsol} should have mass dimension $+1$ and should also transform as 
\begin{equation}
  W(\lambda_0) \to e^{-\Sigma(\lambda_0)} W(\lambda_0)\; ,
\end{equation}
to ensure that the open Wilson line has a covariant transformation law
\eqref{scale-W}. We also demand that the action \eqref{action-W}
reduces to the pure gauge case \eqref{action-IW1} in the integrable
geometry limit. Setting $b_\mu =\frac1\phi
\partial_\mu \phi$ the functional $W(\lambda)$ becomes
\begin{equation}
  W(\lambda) = \frac{W(\lambda_0)}{\phi(\lambda_0)} \phi(\lambda) \; ,
\end{equation}
so, in this case, up to a true constant $\tfrac{W(\lambda_0)}{\phi(\lambda_0)}$
(which is dimensionless and Weyl invariant), $W(\lambda)$ is given by
the dilaton $\phi(\lambda)$. Hence we obtain that in the pure gauge limit we must have
\begin{equation}
  \label{Wphi}
  W(\lambda_0) \rightarrow \phi(\lambda_0) \; .
\end{equation}
Therefore, the final action in the non-integrable
case can be written as\footnote{One could interpret the factor $W(\lambda)$ as a Wilson line dressing for the metric $g_{\mu \nu}$ which can define a map between Weyl geometry and Riemannian geometry as recently proposed in \cite{GM}.}
\begin{equation}
  \label{action-W1}
  S =  -\xi W(\lambda_0) \int_{\gamma}^{} e^{\int_{\lambda_0}^\lambda
    b(\lambda') d\lambda'} 
  \sqrt{-g_{\mu \nu} \dot x^\mu \dot x^\nu} d \lambda \; .
\end{equation}
Another requirement that the action needs to satisfy, essential for path integral quantization is additivity. Defining the action between two arbitrary points $a$ and $b$ on the path as
\begin{equation}
  S_{ab} =-\xi W(\lambda_a) \int_{\lambda_a}^{\lambda_b}
  e^{\int_{\lambda_a}^\lambda  b(\lambda') d \lambda'}
  \sqrt{-g_{\mu \nu} \dot x^\mu \dot x^\nu} d \lambda \; ,
\end{equation}
it is easy to check that we have
\begin{equation}
S_{ac} = S_{ab} + S_{bc} \; ,
\end{equation}
by using the identity
\begin{equation}
W(\lambda_b) = W(\lambda_a) e^{\int_{\lambda_a}^{\lambda_b} b(\lambda') d\lambda'} \; ,
\end{equation}
which derives from equation \eqref{Wsol}.
Therefore we have defined a consistent action, equation \eqref{action-W1}, which is
dimensionless, Weyl invariant and correctly reproduces the Weyl
geodesic equation. Moreover, in the pure gauge case, it reduces to the
action \eqref{action-IW1} which also reproduces the Riemannian action
once we break the scale symmetry. Conditions for finding an action for
the geodesic equation written with a non-metric connection were given
in \cite{LH} and it is not hard to check that our setup does indeed
satisfy them.

To our knowledge, this action was not written before in the literature and it represents
 a central point of our paper. Parts of it were obtained before
\cite{Romero2,Romero3,Quiros,LH}, but the accent was put on finding a
definition for the proper time and the second clock effect.
On the contrary, being 
dimensionless, this action can not be used to measure the proper time,
while removing the factor  $W(\lambda_0)$ in front, in order to obtain the correct time (inverse mass) dimension, is not an option
as the action becomes non-additive. Therefore, in the unbroken phase of the theory, the action which reproduces the geodesic equation in Weyl geometry -- and therefore defines ``straight paths'' in a Weyl sense -- can not give us a proper time as it was the case in Riemannian geometry. This will however be possible in the broken phase as we show below.

\subsection{Spontaneous breaking of Weyl symmetry}

We saw before that breaking the Weyl symmetry in the integrable case
is quite intuitive. We want to do the same on the general case where
the Weyl field is dynamical and becomes massive after symmetry breaking
\begin{equation}
  b_\mu \to B_\mu = b_\mu - \frac1\phi \partial_\mu \phi \equiv -
  \frac1\phi \hat \nabla_\mu \phi \; ,
\end{equation}
where $B_\mu$ is the massive Weyl gauge field which absorbs $\phi$ via a Stueckelberg mechanism. In terms of it, the open Wilson line $W(\lambda)$ becomes
\begin{equation}
W(\lambda) = \frac{W(\lambda_0)}{\phi(\lambda_0)} \phi(\lambda) e^{-\int_{\lambda_0}^{\lambda} B(\lambda')d\lambda'} \; .
\end{equation}
If one defines $\Omega(\lambda)$ as
\begin{equation}
  \Omega(\lambda) \equiv \int_{\lambda_0}^\lambda B(\lambda')d\lambda' =
  - \int_{\lambda_0}^\lambda \frac1\phi \frac{d \phi}{d
    \lambda'} d \lambda' + \int_{\lambda_0}^\lambda b(\lambda')
  d \lambda' =   
  -\ln{\frac{\phi(\lambda)}{\phi(\lambda_0)}} +
   \int_{\lambda_0}^\lambda b(\lambda')d\lambda' \; ,
  \end{equation}
then the particle action of Weyl gravity can be written as
\begin{equation}  
  S= -\xi \frac{W(\lambda_0)}{\phi(\lambda_0)} \int_\gamma \phi(\lambda)
  e^{\Omega(\lambda)}  \sqrt{-g_{\mu \nu} \dot x^\mu \dot x^\nu} d \lambda \; .
\end{equation}
Setting $\phi$ to its vacuum expectation value, $\langle \phi
\rangle$, and decoupling the
massive Weyl field $B_\mu \to 0$ (or equivalently
$\exp{\Omega(\lambda)} \to 1$), we find again the Riemannian
world-line action with the mass parameter given by $\xi \langle \phi
\rangle$. Indeed, decoupling $B_\mu$
effectively imposes the pure gauge condition $b_\mu \rightarrow
\frac1\phi \partial_\mu\phi$ and therefore we also have $W(\lambda_0)
\rightarrow \phi(\lambda_0)$ from equation \eqref{Wphi}. If fluctuations around the vacuum
  are considered we may obtain Weyl geometry induced corrections to the
  Riemannian action and geodesic equation. Studying such corrections
  is not the purpose of 
  this paper, but on general grounds we expect that they are
  suppressed by the symmetry breaking scale.
Needless to say, in the broken phase, where the mass
  parameter $m=\xi \langle \phi \rangle$ is generated, the action can
  be again used to define the proper time. The corrections to the
  Riemannian action mentioned above have the potential to induce a
  second clock effect on particles which couple directly to the
  dilaton, in particular the Higgs field.

\subsection{Affine parametrizations and proper time}

So far we only discussed the possibility of defining a world-line
action from which the geodesic equation for Weyl geometry can be
derived. Unlike the Riemannian case this action can not be used in
order to measure a proper time, as the action is dimensionless.
The question then remains, can we still define a proper time in the
unbroken phase of Weyl geometry? 
Building on \cite{EPS,Perlick,Romero2} and on the general discussion of GR, we consider as a definition for the proper time the quantity, $\tau$, which has inverse mass dimension and provides an affine parametrization of the geodesic equation.
Moreover, in Weyl geometry observables have to be scale gauge
invariant and therefore $\tau$ should have zero scaling dimension
and obviously, it should also be additive. As we shall see these four
requirements can not be satisfied simultaneously.

Let us first focus on finding
the affine parametrizations of the gauge covariant derivative $\hat
\nabla$. We consider the geodesic
equation in Weyl geometry, equation \eqref{Wgeo-u}, and rewrite it as in GR
\begin{equation}
  \dot x^\mu
  \hat \nabla_\mu \dot x^\rho +\frac12
  \dot x^\mu \dot x^\rho \frac{\hat \nabla_\mu(-g_{\gamma\delta} \dot
  x^\gamma \dot x^\delta)}{g_{\alpha
          \beta}\dot x^\alpha \dot x^\beta} =0 \; .
\end{equation}
In the second term, under the derivative there is a scalar combination,
but which has Weyl charge $2(p+1)$ and using equation \eqref{action-gauge} we obtain
\begin{equation}
  \dot x^\mu \hat \nabla_\mu \dot x^\rho - \frac12 \dot x^\rho
    \dot x^\mu \left [\partial_\mu\ln{(-g_{\alpha\beta} \dot  x^\alpha \dot
          x^\beta)} + (2p+1)b_\mu \right] =0 \; .
\end{equation}
This can be written in the form of a pregeodesic
\begin{equation}
   \dot x^\mu \hat \nabla_\mu \dot x^\rho - \frac12 \dot x^\rho
   \frac{d}{d \lambda} \ln \left (- W(\lambda)^{2(p+1)} g_{\mu \nu}
     \dot x^\mu \dot x^\nu \right) =0 \; ,
\end{equation}
with the help of equation \eqref{eq-W}.
The argument of the logarithm is Weyl invariant and can be
consistently set to 1 to obtain the affinely parametrized geodesic equation in Weyl
geometry. Therefore the generalization of the condition \eqref{prtcond} to Weyl geometry is
\begin{equation}
W^{2(p+1)} g_{\mu \nu} \frac{d x^\mu}{d \tilde \tau} \frac{d x^\nu}{d \tilde \tau} = -1\; .
\label{affine-Weyl}
\end{equation}
Consistency of this equation requires that (see also table \ref{dimensions})
\begin{align}
[d \tilde \tau]_W = - p\; , && [d \tilde \tau]_M = M^p\; ,
\end{align}
since the LHS of equation \eqref{affine-Weyl} must also have zero mass
dimension. As it is apparent from above there is a tension between
having dimension of time (inverse mass) and being a gauge invariant
observable. Rescaling $\tilde \tau$ by a dimensionful parameter is
not an option in the symmetric phase of the theory. If one rescales by
a compensator field, like the dilaton, then the affine
parametrization is lost. We conclude that there is no definition
of proper time in Weyl geometry which has the correct inverse
  mass dimension, is Weyl invariant and additive and also gives an
  affine parametrization. However, after scale symmetry breaking,
which essentially sets $W=\langle \phi \rangle$, 
with 
  an appropriate rescaling by the {\it vev} of $\phi$, namely $\tau = \langle \phi \rangle ^{-(p+1)} \tilde \tau$, one can recover
  the usual notion of proper time. 

We have used so far an arbitrary charge $p$ for the tangent vector
$\dot x^\mu$ of the curve followed by the particle. The
geodesic equation in an arbitrary parametrization could 
be written in terms of the normed vector $u^\mu$ which always has
charge $-1$, thus independent of the charge
$p$. Therefore 
different charges yield the same (pre)geodesics in the sense of
belonging to the same projective structure as we discussed in the
GR case. This can be seen by the fact that the terms
depending on $p$ are of the form $\dot x^\mu b_\mu \dot x^\rho \sim f
\dot x^\rho$.

 However, the charge $p$ does
play a role in finding an affine parametrization, namely the
gauge invariant condition \eqref{affine-Weyl}. 
Two cases seem to be special when looking at this equation. First,
choosing $p=-1$, \cite{HL1,GM,GM-Ligo}
corresponds to geometric vectors (their charge is
given exclusively by the vielbein)
for which, the gauge covariant derivative
  $\hat \nabla$ on $\dot x^\mu$ is the same 
  as the affine derivative with torsion $\nabla$. Equation
\eqref{affine-Weyl} reduces formally to the GR
condition $g_{\mu \nu} \frac{d x^\mu}{d \tilde \tau} \frac{d x^\nu}{d
  \tilde \tau} = -1$ and therefore non-locality is absent. The quantity $\tilde \tau$ defined like this has the right
 dimension of time, but it is not gauge invariant and therefore not an
 observable. Rescaling it with a compensator field (which needs
 to be dimensionless!) would spoil the affine parametrization.
 Another special case is that of invariant tangent vector
 corresponding to charge $p=0$, \cite{Romero2}. Now, the gauge covariant
 derivative 
 $\hat \nabla$ coincides with the
 non-metric affine derivative $\tilde \nabla$. The parameter
 $\tilde \tau$ satisfies $ W^2 g_{\mu \nu} \frac{d x^\mu}{d \tilde
   \tau} \frac{d x^\nu}{d \tilde \tau} = -1$ and it is both
 dimensionless and gauge invariant. Actually, only for this charge, $\tilde \tau$ is equal to the
 action \eqref{action-W} up to the parameter $\xi$. Instead, for a general charge we have
\begin{equation}
  \tilde \tau = \int W^{p+1} \sqrt{-g_{\mu \nu} \dot x^\mu \dot x^\nu} d\lambda\; .
\end{equation} 
  Before scale
 symmetry breaking there is no dimensionful parameter to rescale it
 in order to yield the right mass dimension of time. Rescaling it by a
 field (of mass dimension $+1$ and Weyl weight $0$) would spoil the
 affine parametrization. Dividing by $W(\lambda_0)$ would both spoil
 the gauge invariance and additivity. We conclude that no value of the
 charge $p$ can give us a satisfactory notion of proper time in the
 unbroken phase of the theory.
 Therefore, if any second clock effect
exists this should necessarily appear in the broken phase and
  should be traced in the Weyl-induced corrections to Riemannian geometry.

Finally, let us shortly discuss the quadratic action which may be
relevant for path integrals. Recall the quadratic action from GR, equation \eqref{Se}, where we introduced an einbein which may be
eliminated from its algebraic equation of motion to recover the
original square root action \eqref{GRaction}. The situation is very
similar in Weyl geometry with the difference that appropriate powers of $W(\lambda)$ are inserted to render equation \eqref{Se} invariant. Let us assume an arbitrary Weyl charge for the einbein $[\e]_W = q$ then the action reads
\begin{equation}
  S= \frac12 \int_\gamma \left [\e^{-1}W(\lambda)^{(2+p-q)} g_{\mu \nu} \dot
    x^\mu \dot x^\nu -\xi^2 \e W(\lambda)^{q-p}
  \right] d \lambda\; .
\end{equation}
By using the equation of motion of the einbein
\begin{align}
\frac{\dot x^2}{\e^2} + W^{2(q-p-1)}\xi^2 = 0\; , && \implies && \e=\frac{\sqrt{-\dot x^2}}{\xi} W^{p-q+1} \; ,
\label{einbein-eom}
\end{align}
one can easily check that we recover equation \eqref{action-W}.
In Riemannian geometry, the second term of the action was a constant and could be discarded as not contributing to the equations of motion, but here, this is not the case and it gives a distinct
contribution to the equations of motion which read
\begin{equation}
\dot x^\mu \hat \nabla_\mu \dot x^\rho - \frac{d}{d \lambda} \ln (W^q \e)\dot x^\rho = 0\; .
\end{equation}
Details of the calculation are similar
to what has been presented so far. As before, the pregeodesic term contains the Weyl invariant combination $W^q \e$ which can be set to be equal to a constant. The analogous gauge condition \eqref{gauge-e} for Weyl geometry is
\begin{equation}
W^q \e = \frac{1}{\xi^2} \;.
\end{equation}
Combining the above with the equation of motion of the einbein \eqref{einbein-eom} yields the condition for an affine parametrization in the form
\begin{equation}
W^{2(p+1)} g_{\mu \nu} \frac{d x^\mu}{d \tilde \tau} \frac{d x^\nu}{d \tilde \tau} = -\frac{1}{\xi^2}\; ,
\end{equation}
analogous to the GR condition 
\eqref{prtcond}. A further rescaling of $\tilde \tau$ with the (inverse of the) dimensionless parameter $\xi$ recovers the previously found condition \eqref{affine-Weyl}. Hence in Weyl geometry as well one can successfully eliminate the square root from the action. However, the non-local Wilson line factors may be concerning for the purpose of path integral quantization. In this respect one can treat $W(\lambda)$ as a constrained field on the world-line satisfying the covariant constancy condition $\hat D_\lambda W = 0$ implemented with the help of a Lagrange multiplier $\eta$ with appropriate Weyl weight
 \begin{equation}
  S= \frac12 \int_\gamma \left [\e^{-1}W^{(2+p-q)} g_{\mu \nu} \dot
    x^\mu \dot x^\nu -\xi^2 \e W^{q-p}
   + \eta \hat D_\lambda W \right] d \lambda\; .
\end{equation}  
The action above is scale invariant, additive, dimensionless and local. It can therefore constitute a good starting point for path integral quantization. The non-locality we encountered earlier is then simply the result of integrating out the constrained field $W(\lambda)$ from the action above.

\section{Conclusions}

In this paper we constructed world-line actions for a particle
moving on a time-like curve in Weyl geometry. Our starting point was the modern gauge theory definition of Weyl gravity. As such we worked with tangent vectors to the curve with arbitrary Weyl charges. 
Compared to  Riemannian
geometry, finding a world-line action poses some new challenges. Since no mass parameter is
allowed by the symmetry, writing a dimensionless functional needs
additional fields (aside from the metric) in the theory. One choice is the dilaton,
defined in equation \eqref{phidef}, the scalar mode propagated by the $R^2$ term contained in the spacetime action. We show that this leads to integrable Weyl
geometry. The way to find an action in the non-integrable case is to
add a non-local (path dependent) factor \eqref{Wsol}. Then, a
consistent action, which is dimensionless, Weyl invariant and additive
can be defined. In the symmetric phase this action can not define a
proper time as no mass parameter exists. Breaking the symmetry
however, induces a mass proportional to the \textit{vev} of $\phi$,
the action in the vacuum becomes Riemannian and the notion of proper
time again makes sense.
We also consider a more systematic way of defining the proper time by considering affine parametrizations of the time-like curve,  as we actually do in ordinary GR.
In the unbroken phase of the theory this again does not lead to a
satisfactory notion of proper time. More precisely, we find that asking the
proper time to have inverse mass dimension, to be additive and Weyl
invariant together with the condition that it gives an affine
parametrization can not be satisfied at the same time. One may question whether all four conditions we require are actually necessary. From this point of view
it is not clear that our construction exhausts all the possibilities to define the proper time. Nevertheless, this is a strong result which confirms the physical intuition that in a scale-invariant theory it does not make much sense to consider dimensionful observables.

The action we construct was actually used in the literature as
defining the proper time \cite{Romero2, HL1}. The presence of a history
dependent term in this action, $W(\lambda)$, has fueled a whole debate
on the existence of a second clock effect in Weyl geometry. The point
we advocate is that since our action does not have the correct inverse
mass dimension in order to be associated to a time, a proper time in
the unbroken phase of the theory can not be defined and no second
clock effect can be claimed to exist as long as the scale symmetry is
intact. Breaking this symmetry leads to a Riemannian action, but
Weyl-induced corrections have the potential to generate such an effect
in the broken phase. However, in the broken phase the Weyl gauge field becomes massive and interactions are short ranged. Moreover, such an effect can only be observed on
particles which directly couple to the dilaton, in particular the
Higgs boson and if indeed present we expect it to be suppressed by the
symmetry breaking scale.
Studying the corrections on Einstein gravity induced by
Weyl geometry is an interesting subject on which we hope to come back
soon \cite{CM2}.

Aside from the gauge covariant formulation of Weyl geometry used throughout the paper there exist two other non gauge covariant formulations described by affine connections with (vectorial) non-metricity and torsion respectively. We point out that the geodesics (or rather the autoparallels) of the corresponding (affine and gauge) covariant derivatives are the same as they belong to the same projective structure (though their affine parametrizations are different).

Finally, we constructed an equivalent quadratic action by introducing an einbein on the world-line and corresponding powers of the non-local factor $W(\lambda)$ in order to render it Weyl invariant. Such an action, free of square roots, is more amenable for path integral quantization. Moreover, a fully local version can be written by treating $W(\lambda)$ as a world-line field constrained to satisfy the covariant constancy equation by a Lagrange multiplier. It is then apparent that non-locality arises only when one integrates out this field.\\[.5cm]
\noindent
{\bf Acknowledgments: } This work was supported by the Ministry of Research, Innovation and Digitization, CNCS-UEFISCDI, project number PN-23210101/2023.

\end{document}